\documentclass[12pt]{iopart}
\eqnobysec
\usepackage{graphicx}
\usepackage{color}
\usepackage{subcaption}
\usepackage{bm}

\font\amsb=msbm10
\def\hbar{\mbox{\amsb\char'175}}

\newcommand{\be}{\begin{equation}}
\newcommand{\ee}{\end{equation}}

\newcommand{\vecp}{{\mathbf p}}
\newcommand{\vecq}{{\mathbf q}}

\newcommand{\x}{{\mathbf x}}

\newcommand{\der}{\partial}

\begin{document}

\title{Are quantum trajectories suitable for semiclassical approximations?}

\author {Alfredo M. Ozorio de Almeida\footnote{alfredozorio@gmail.com}}
\address{Centro Brasileiro de Pesquisas Fisicas,
Rua Xavier Sigaud 150, 22290-180, Rio de Janeiro, R.J., Brazil}

\begin{abstract}

The quantum trajectories in the de Broglie - Bohm formulation of quantum mechanics depend on an additional quantum potential derived from the full wave solution of Schr\"odinger's equation. The task of supplying collectively all the correct quantum results strongly alters the characteristics of the corresponding classical trajectories, which underlie semiclassical approximations to the evolving wave function. Both classical and quantum trajectories are here considered to be conservative with no influence of an external environment, even though this is the source of eventual classicality in quantum systems, that is, decoherence.

The concept of integrability, closely correspondent in classical and quantum mechanics, is not preserved by the quantum trajectories. General systems, in which classical chaotic motion participates, are much harder to treat semiclassically, but quantum trajectories can be chaotic even for integrable systems. This discrepancy between the character of classical and quantum trajectories in the de Broglie - Bohm  interpretation does not clarify the singular classical-quantum transition. 

\end{abstract}

\maketitle

\section{Introduction}

Initial reactions to David Bohm's bold reinterpretation of quantum mechanics in terms of
trajectories, which are sensitive to the evolving wave function \cite{Bohm52a, Bohm52b}, were mostly negative.
Even though this was preceded by De Broglie's work on pilot waves \cite{debrog}, 
the conceptual immobility imposed by the dominant Copenhagen interpretation was broken
in the long run by Bohm showing that quantum trajectories are fully compatible with standard quantum mechanics. 
Indeed, the recent meeting "Bohm in Brazil", held at the Advanced Study Institute
of the University of Sao Paulo, brought to light how this influenced the proposal by John Bell 
of the fundamental experiments \cite{Bell,speakable}, which have established a more solid ground for our conceptions.

Quantum mechanics (QM) covers vast ranges, from elementary spins or q-bits to those manybody systems, 
which may have properties that are accurately described classically.
In its centenary, many different approaches and representations have shown their value
in treating a growing list of quantum problems and applications. My concern here is with semiclassical aproximations
(SC), that is, with the supplementation of classical mechanics (CM) that is required to treat quantum problems
with increasing accuracy as Planck's constant $\hbar \rightarrow 0$, compared to appropriate classical actions.

I will not deal with this classical limit itself, which has a remarkable singular nature 
discussed, for instance, by Michael Berry in \cite{BerHouches}. Therefore, no matter how close to the classical limit, 
it is QM that is here considered.
Indeed, this is unitary QM, previous to any measurement and sealed
against any influence by an environment or heat bath, which drives
the transition to classicality within the various models of decoherence \cite{Giulini}. The corresponding classical trajectories for such an open quantum system need to be dissipative \cite{Aldo}, instead of the conservative classical and quantum trajectories discussed here. 

The starling conceptual break from CM to QM forms the initial barrier to the development of SC approximations. At first sight,
the uncertainty principle seems to bury the possible relevance of any kind of trajectory initiated at a single point in phase space. SC approximations are thus a direct growth of the old quantum theory of Bohr \cite{Bohr} and Sommerfeld \cite{Somm}, rather than germinating within the main stream of QM. The neatest connection
to bona fide QM proceeds by letting the classical variational principle pick out the trajectories, 
together with their phases, in Feynman's path integral \cite{Feynman, Schulman,Morette}.

Nevertheless, wouldn't it be more satisfying to start from the de Broglie-Bohm alternative, 
even though also exact version of QM?  
Each experimental result then depends on a particular quantum trajectory, within a probability ensemble that
evolves in time along with Schr\"odinger's wave function. At a first glance, the approximation of a quantum trajectory
by its classical counterpart should be far simpler than approximating entire quantum states. The difficulty lies in
Bohm's quantum potential: the trajectories generated with this apparently innocent addition to the classical potential
have the enormous task to supply all the exact results of QM. No classical perturbation theory \cite{Arnold} is able to bridge this gap. For this reason, the vast simplification of first order classical perturbation theory \cite{GOE}, which would merely add the integral of the quantum potential over the unperturbed classical trajectory to obtain its Bohm phase, is doomed to be misleadingly inaccurate. 

A further obstacle arises if there is  more than one degree of freedom. It is well established that integrable systems,
which generalize in a rather straightforward way those with a single degree of freedom, are a minority among systems that harbour
chaotic, exponentially divergent pairs of classical trajectories. Indeed, arbitrarily small perturbations of integrable systems 
already generate narrow chaotic zones. Thus, even though classical and quantum integrable Hamiltonians are often matched, 
along with their respective constants of the motion (in the form of phase space functions or commuting observable operators),
the correspondence relies on purely classical trajectories, unaffected by the quantum potential.

The following section presents the de Broglie-Bohm interpretation and a short review of standard criticism for simple systems.
Then section 3 deals with classical integrability and its quantum correspondence.
The breaking of integrability, leading to classically chaotic trajectories, is introduced in section 4, whereas the SC treatment
of chaos is discussed in section 5. Concluding remarks concerning the difficulty of accessing SC approximations with quantum trajectories
are presented in section 6.

\section{Traveling waves and stationary waves}

For a start, let us take the simplest examples of wave functions of a quantum system, solutions of the Schr\"odinger equation 
in the configuration space $\vecq=(q_1,...q_N)$
\be
\left[-\frac{\hbar^2}{2m}\nabla^2 + V(\vecq) \right ] 
\langle \vecq|\psi \rangle_t 
= i\hbar ~ \frac{\der \langle \vecq|\psi \rangle_t}{\der t}
\ee
and their de Broglie-Bohm trajectories \cite{Bohm52a}. Inserting the complex Bohm ansatz
\be
\langle \vecq|\psi \rangle_t = R_t(\vecq) ~ \exp\left(\frac{i}{\hbar} S_t(\vecq) \right) ~,
\label{Bohmwave}
\ee
into the real part of the Schr\"odinger equation leads to an analogue of the Hamilton-Jacobi equation
\be
-\frac{\der S_t}{\der t} = \frac{\nabla S_t ^2}{2m} + V + Q_t ~,
\ee
with the added time-dependent quantum potential
\be
Q_t(\vecq)= -\frac{\hbar^2 }{2m}\frac{\nabla^2 R_t}{R_t}  ~.
\label{quantumpot}
\ee
Therefore, the quantum trajectories depend on the curvature of the evolving amplitude of the exact wave functions.  

In the case of a single degree of freedom, $N=1$, $(\vecq=q)$ a plane wave with energy $E$ propagates over a constant potential $V$:
\be
\langle q|\psi \rangle = \exp\left(\frac{i}{\hbar} pq \right) ~,
\label{planewave}
\ee
where the constant momentum $p=\sqrt{2m(E-V)}$ (while the normalization constant is dropped). This is already in the standard Bohm form 
\eref{Bohmwave}, here with the linear phase $S(\vecq) = pq$ and constant amplitude $R(\vecq)=1$, which also 
coincides with the simplest SC approximation. Indeed, even for a variable potential with $E>V(q)$ everywhere,
the SC generalization of the traveling plane wave \eref{planewave} has its phase
defined by the classical action
\be
S_{\rm SC}(q)= \int_{q_0}^q p(q') ~ \rm{d} q' = \int_{q_0}^q \sqrt{2m(E-V(q'))} ~ \rm{d} q' ~.
\ee 
The amplitude 
\be
R_{\rm SC}(q)= \frac{1}{\sqrt {p(q)}} ~,
\ee
is the square root of the constant classical Liouville distribution in phase space \cite{Arnold}, 
projected onto the position axis \cite{BerMount}. 
On the other hand, the complete quantum wave function includes a relatively weak reflected wave from even a soft step, 
in which the classical potential is increased by $\delta V$  within a stepwidth $\delta q$, that is accounted for within 
the complete Bohm description. Though incorporated into more refined SC approximations \cite{BerMount}, 
this backscattering vanishes in the full classical limit,
\be
\frac{p \delta q}{\hbar} \rightarrow 0.
\ee

Backscattering comes into full view if there is a turning point $q_0$ where $E=V(q_0)$.
Then, given an incoming wave with unit amplitude, the amplitude $B$ of the backscattered wave is appreciable, that is,
\be
\langle q|\psi \rangle_{SC} = \frac{1}{\sqrt {p(q)}} ~ \left[\exp\left(\frac{i}{\hbar} S_{\rm SC}(q) \right) 
+ B \exp\left(\frac{i}{\hbar} S_{\rm SC}(-q) \right)  \right] ~,
\ee
together with a transmitted wave with amplitude $T$
\be
\langle q|\psi \rangle_{SC} = T \left[\exp\left(-\frac{1}{\hbar}\int_{q_0}^q  \sqrt{2m(V(q')-E)}\rm{d} q'  \right) \right] 
\ee
on the other side of the turning point, which may be attributed to an evanescent transmitted trajectory. Here, the absence of classical  transmission is neatly reproduced in the classical limit. On the classically allowed side, one sees that the simple standing wave
\be
\langle q|\psi \rangle = \left[\exp\left(\frac{i}{\hbar} pq \right) + \exp \left(-\frac{i}{\hbar} pq \right) \right]
= 2 \cos\frac{pq}{\hbar}  ~
\label{standingwave}
\ee
is also reproduced with $p=\sqrt{2m(E-V)}$, in the case of a constant potential $V$ on one side of a sharp potential step.

In contrast, given its general form \eref{quantumpot},
the quantum potential for quantum trajectories reduces in this case to
\be
Q(q) = \frac{p^2}{2m}= E-V,
\ee
which freezes all the quantum orbits! 
\footnote{Indeed, it is easy to show that this curious static result holds generally for Hamiltonian eigenstates
and arbitrary stationary quantum states (see e.g.  \cite{Poirier1}). With hindsight, this makes sense, since the Bohm ansatz is modeled on a traveling wave.}
Thus, the superposition of two waves traveling in opposite directions, which classically corresponds to an ensemble of particles 
traveling in both directions, is reduced to immobility for Bohm's quantum trajectories. Notwithstanding its correct description
of the quantum scenario, there is no smooth classical limit as $\hbar \rightarrow 0$.

\section{Integrability for more degrees of freedom}

The classical phase space of a system with $N$ degrees of freedom adds a further $N$ momenta $\vecp=(p_1,...,p_N)$
to the positions $\vecq$, so that its points are $\x=(\vecq, \vecp)$. Following Arnold \cite{Arnold}, a classical Hamiltonian $H(\x)$
is integrable if there exist $N$ independent functions $F_n(\x)$ (including the Hamiltonian), such that the Poisson Brackets
\be
\{F_n(\x),F_{n'}(\x))\} = \sum_{m=1}^N \left(\frac{\der F_n}{\der p_m} \frac{\der F_{n'}}{\der q_m} 
- \frac{\der F_n}{\der q_m} \frac{\der F_{n'}}{\der p_m} \right)(\x) =0 ~,
\ee
for all $n$ and $n'$. 
\footnote{Such a set of phase space functions are said to be in involution \cite{Arnold}.
The classical Hamiltonian is limited to the form $\frac{\vecp^2}{2m} + V(\vecq)$ so as to correspond to Schr\"odinger's equation,
but the other phase space functions are not constrained.}
This not only implies that each of these functions is a constant along all the classical trajectories
generated by Hamilton's equations, but that they would still be constant, were any other of them singled out as an alternative 
Hamiltonian: $F_n(\x) = H'(\x)$. 
The construction of such complete sets of constants of the motion is at the heart
of powerful methods of solving Hamilton's equations and most physicists, including the founders of QM, were not aware
that Poincar\'e had proved that generally classical systems are not integrable \cite{Poincare}. 

Correspondence of the classical Poisson bracket with the commutator between observable operators was one of the pillars for
the early development of QM. Then, it is no surprise that the quantum version of integrability 
requires the existence of $N$ observable operators ${\hat F}_n$ (including the Hamiltonian $\hat H$), which commute among themselves:
\be
[{\hat F}_n ,{\hat F}_{n'}] = {\hat F}_n {\hat F}_{n'} - {\hat F}_{n'} {\hat F}_n = 0 ~.
\ee
Heisenberg's equations of motion then imply that anyone of these operators remains constant under the action of each of the others,
should it be chosen as the Hamiltonian.

Even though quantum integrability is just as rare as its classical correspondent, the old quantum theory of Bohr-Sommerfeld
was built on it and so are most of the SC approximations of full QM. In any case, the correspondence of classical constants of the motion
$F_n(\x)$ to quantum constant observables ${\hat F}_n$ and hence the constancy of the expectation $\langle{\hat F}_n\rangle$ of any measurement 
relies on classical trajectories. 
The classical trajectories of an integrable system are constrained to lie in the intersection of the $N$ surfaces $F_n(\x)=C_n$,
that is, an $N$-dimensional surface in the $2N$-dimensional phase space. However, there is no such constraint for the corresponding quantum trajectories, which depend on the generally time dependent quantum potential, thus breaking integrability.

\section{Chaotic trajectories in phase space}   

Weak perturbations of a classical integrable Hamiltonian generably break integrability, that is,
there is generally not even a single other constant of the motion, let alone $N$ constants in involution.
With increasing perturbation, wider and wider regions exhibit the exponential divergence of neighbouring
trajectories, which characterizes classical chaotic behaviour.
The discovery of chaotic classical motion was overshadowed in the physics community by the breakthroughs of quantum mechanics.
Nevertheless, the inexistence of a direct quantum correspondence, for the startling complexity of general classical motion
that is clearly observable in nature, has even led to the questioning as to which theory is more fundamental \cite{FordMant}.
  
The break from integrability does not impede the construction of SC approximations for the evolution of wave functions
for a finite time. Starting with the work of Van Vleck (1928) \cite{VanVleck,Gutzbook}, one can follow an initial state constructed semiclassically on an appropriate phase space surface, until its chaotic evolution crinkles the surface to an unmanageable degree. The full difficulty arises for stationary properties, such as the eigenstates of chaotic Hamiltonians or just the density of eigenstates: to these an infinite evolution time may be ascribed. For weak perturbations, an approximation by integrable systems still renders acceptable SC approximations \cite{LanAlm}, since isolated $N$-dimensional surfaces survive, which limit breakaway trajectories according to the KAM theorem \cite{Arnold}.  

The development of SC methods to deal with chaotic classical motion is discussed in the following section.
Here it must be observed that, once again, quantum trajectories are of no help. Indeed, the dependence of
the quantum potential on the initial state forces it to be likewise time dependent, if the wave function is not invariant.
But  the addition of a time-dependent quantum potential generally destroys the integrability of a static classical Hamiltonian,
so that the guiding structure of invariant $N$-dimensional surfaces for the SC approximatons is lost.
In other words, the quantum trajectories wash away the safe niche of integrable systems, even if they are rare.

Let us consider, for instance, an initially localized wave packet inside a square quantum billiard, i. e, a square with a constant potential
and hard boundaries. This system is integrable, since the momentum operators ${\hat p}_1$ and ${\hat p}_2$ commute with the
Hamiltonian ${\hat H}= {\hat p}_1^2/(2m) + {\hat p}_2^2/(2m)$ inside the square, and the classical momenta are likewise 
constants of the motion. In contrast, the amplitude in the Bohm wave \eref{Bohmwave} will have a time-dependence
that is imparted to the quantum potential \eref{quantumpot}. As described in  \cite{Matzkin}, the trajectories are no longer made up of straight line segments and the integrable constraints of the classical motion are broken. 
The quantum potential is not even periodic, since the wave function will become evermore crinkly as the wave packet broadens 
and multiply reflects on the sides of the square.
It follows that the clarity of the parallel between classical and quantum constants of the motion is in no way propitiated
by the quantum trajectories.

 \section{Semiclassical approximations for chaotic systems}

Frustration in the search for acceptable SC approximations of stationary quantum features of classically chaotic systems
seems to be imposed by the complexity of nonintegrable classical motion. The procedure to circumvent this conundrum is inspired
by Poincar\'e's observation that simple selfrepeating periodic orbits with arbitrarily long periods are dense in all Hamiltonian systems
with bounded motion. This suggests that one should keep track of the dense set of periodic orbits that survive any perturbation of the Hamiltonian.

The earliest SC instance of this approach was implemented for fully chaotic systems, in which all periodic orbits are unstable, 
by Gutzwiller \cite{Gutzbook}, concurrently with Balian and Bloch \cite{BalBloch}.
A contemporary expression of their results is that the smoothed density of eigenstates of a chaotic Hamiltonian with energy $E_k$, 
\be
\delta_\epsilon(E) \equiv \sum_k \frac{1}{\pi} ~ \frac{\epsilon}{\epsilon^2 + (E-E_k)^2} ~,
\label{widelta}
\ee
can be separated into two terms. One is a classical term, determined by the volume of the smoothed energy shell $H(\x)=E$, 
so that it is smoothly energy dependent (see e.g.\cite{BerHouches}). 
The oscillatory quantum dependence is a sum, with each term contributed by a periodic orbit. 
Though isolated in the energy shell $H(\x)=E$, their classical action depends on the energy and determines the phase of the contribution. 
Thus these periodic orbit terms oscillate in energy: the longer the orbit, the faster the oscillation. 
On the other hand, the energy window provides a cutoff for the very long periodic orbits: for $\hbar/\epsilon$ smaller
than the period of the shortest periodic orbit, all oscillatory terms are canceled and only the classical term is left
in the ultra-smoothed density of states. 

For the SC treatment of manybody systems with very high densities of state, it may be hard to distinguish a single eigenstate, specially as chaos
breaks any selection rule that would help to pinpoint it. Even so, one would hope to squeeze somehow the energy window sufficiently
so as to find individual eigenstates. This is indeed achieved by various schemes of ressumation of the periodic orbits,
though at the cost of increased complexity of the SC approximations \cite{Haake, Bog90,BerKea}.
Notwithstanding the satisfaction of finding how it all works out,
it is doubtful if a practical SC method is available for calculating the unsmoothed density of states.  

Various SC approximations of the density operator composed of mixed eigenstates, in a narrow energy window
\be
\rho(E|\epsilon) \equiv \sum_k \frac{1}{\pi} ~ \frac{\epsilon}{\epsilon^2 + (E-E_k)^2} |E_k\rangle \langle E_k| ~,
\label{widestate}
\ee 
are constructed on open orbit segments within the classical energy shell $H(\x)=E$. In other words, this is the density operator
for the coarsegrained microcanonical ensemble.
The choice of segments depends on the 
representation, such as position \cite{Bogscars}, momentum or the Weyl-Wigner representation in phase space \cite{Ber89,Ber89b,Report}. 
Again, the width of the energy window limits the duration of the segments in the SC sum 
and the very short segments require special treatment. The previous consideration about finding individual chaotic eigenstates
in manybody systems holds here, but ressumation methods are also available \cite{Sieber,AgFish}. Curiously they combine periodic
orbits with the open orbit segments. 

More recent is the construction of a SC sum for the probability of a coarsegrained transition between a pair of energy levels $E_k \rightarrow E_j$, driven by an external time-dependent Hamiltonian \cite{Alm24}. Once more, there is a classical term, here determined by the intersection 
of the outgoing energy shell $H(\x)= E_k$ with the classically driven target shell $H(\x(t))=E_j$. 
The quantum oscillatory terms then replace the contributions from periodic orbits in the static density of states 
by those of closed compound orbits, which straddle both energy shells. They are constructed by a pair of segments, one on each shell, 
with their tips joined by two trajectory segments generated by the driving Hamiltonian. 

No previous report of closed compound orbits is known to the author, but their existence in continuous families,
parametrized by the pair of energies and the time in action of the driving Hamiltonian, is easily established \cite{Alm24}. 
A remarkable feature is that these families grow from seeds of instantaneous transitions, in the limit as $E_j \rightarrow E_k$,
located at isolated points where the Poisson bracket of the driving Hamiltonian with $H(\x)$ cancels.
Thus, whereas the full cancelation of the Poisson bracket everywhere reveals a constant of the motion,
its isolated zeroes spark off SC energy transitions, even though no quantum correspondence is available for 
such local features of the Poisson bracket.

\section{Conclusion}

An indirect way that semiclassical methods are able to dig into the messy tangle of chaotic trajectories,
so as to supply valuable information concerning corresponding quantum systems, was briefly reviewed in the previous section.
In all cases the trajectories are purely classical, whether open or periodic, just as in the more standard SC methods
for integrable systems, which grew from the old quantum theory. The setting is classical phase space, which enjoys a flexibility
that is only partially emulated by full quantum systems. Indeed, the invariance with respect to canonical transformations
can only be matched by quantum systems if they are linear, that is, by metapletic quantum transformations \cite{GuilStern,Voros76,AlmIng}.
These transformations can undo the separation of common Hamiltonians into a quadratic function of the momenta and a potential
function of positions, corresponding to the second order Schr\"odinger differential equation. Generalized path integrals in phase space
are not restricted to this simple form of the Hamiltonian \cite{Report} and the classical trajectories, which are singled out by stationary phase integration, remain the basis of SC methods within this wider picture.

Previous versions of the criticism of the De Broglie-Bohm interpretation are reviewed here in section 2. They
have been surmounted by modifications of the Bohm ansatz \eref{Bohmwave}, which add a second wave with its amplitude and phase,
in line with the SC approximation \cite{Poirier1}. This avoids the unintuitive zero-motion trajectories and leads to a smooth classical limit.
Further generalization for integrable systems with higher degrees of freedom is also possible \cite{Poirier2}, 
even though it requires foreknowledge of the classical motion, so as
to include new Bohm waves for each sheet of the invariant surface. The challenge of general classical systems, where chaotic regions
are closely intermingled with invariant surfaces, is perhaps outside the scope of any doctoring.

In the end, the real problem is that the quantum potential, which has been assumed to be known in this discussion, depends on the solution of the Schr\"odinger equation, in its turn the object of SC approximations. Perhaps, the real question should concern the precision of the quantum potential based on a SC approximation to the evolving wave function, thus avoiding the need to integrate a partial differential equation in many variables (at least for nearly integrable systems). First order classical perturbation theory would only supply a correction to the SC phase of an unperturbed (classical) trajectory as in \cite{GOE}, rather than the sought for deviation of the perturbed (quantum) trajectory. Thus the SC version of the quantum trajectory would also be generated by Hamilton's equations, but perturbed by a SC quantum potential. In any case, generally
the quantum potential and hence the effective classical Hamiltonian has a time dependence that depends on the initial quantum state.  

Quantum mechanics is so conceptually rich that a flexibility of representations, such as position, momentum or phase space, wave solutions
of the Schr\"odinger equation, operators evolving according to the Heisenberg equation, as well as interpretations, 
whether Copenhagen or de Broglie - Bohm, all contribute partial glimpses to our understanding and wonder. 
Conversely, each of these different approaches has its blind spots and quantum trajectories have so far not iluminated semiclassical approximations.

\section*{Acknowledgements}
This article is meant to clairfy an implicit scientific point concerning a majour scientific theory. 
Therfore, it should neither compete, nor conflict with the interest of any organization or of any individual.

This work was supported by CNPq [Brazilian Government Agency] under Grant [number 304343/2021-8].
I thank Raul Vallejos for stimulating discussions.

\end{document}